\documentclass[12pt]{iopart}
\usepackage{iopams} 
\usepackage{graphicx}
\begin{document}

\title[Water Absorption on $\mathrm{CeO}_{2}$ at Low Temperatures for Understanding Anti-Icing]{Initial Stages of Water Absorption on $\mathbf{CeO}_{2}$ Surfaces at Very Low Temperatures for Understanding Anti-Icing Coatings}

\author{A C Åsland$^{1}$, S P Cooil$^{2}$, D Mamedov$^{3}$, H I Røst$^{1,4}$, J Bakkelund$^{1}$, Z Li$^{5}$, S Karazhanov$^{3}$ and J W Wells$^{1,2}$} 

\address{$^{1}$ Department of Physics, Norwegian University of Science and Technology (NTNU), NO-7491 Trondheim, Norway.}
\address{$^{2}$ Centre for Materials Science and Nanotechnology, University of Oslo, NO-0318 Oslo, Norway.}
\address{$^{3}$ Department for Solar Energy Materials and Technologies, Institute for Energy Technology, NO-2027 Kjeller, Norway.} 
\address{$^{4}$ Department of Physics and Technology, University of Bergen, NO-5007 Bergen, Norway.}
\address{$^{5}$ Department of Physics and Astronomy - Centre for Storage Ring Facilities (ISA), Aarhus University, DK-8000 Aarhus C, Denmark.} 

\ead{j.w.wells@fys.uio.no}
\vspace{10pt}
\begin{indented}
\item[]\today
\end{indented}

\begin{abstract}
Anti-icing coatings are intended to prevent ice formation on surfaces, minimising the risk of surface-related damage and also reducing ice-related hazards in society.   
$\mathrm{CeO}_{2}$ coatings are robust, hydrophobic, and transmit light, thus they are suitable for a range of applications. 
However, their evolving surface chemistry during the initial stages of $\mathrm{H}_{2}\mathrm{O}$ exposure at very low temperatures has not been investigated, despite that this is important for understanding their anti-icing properties.
  
To study this, $\mathrm{CeO}_{2}$ coatings were grown by sputter deposition, cooled to $\approx100\,$K and exposed to a $\mathrm{H}_{2}\mathrm{O}$ atmosphere at  $1\times10^{-8}\,\mathrm{mbar}$.  

We demonstrate the usefulness of X-ray photoelectron spectroscopy (XPS) as a tool for investigating the anti-icing properties of surfaces. We present XPS measurements of $\mathrm{CeO}_{2}$ coatings before and after exposure to $\mathrm{H}_{2}\mathrm{O}$, in-situ and at cryogenic temperatures.     

XPS reveals that little to no ice forms on the surface of  $\mathrm{CeO}_{2}$ after the $\mathrm{H}_{2}\mathrm{O}$ exposure at $\approx100\,$K. In contrast, ice was observed all over the sample holder on which the $\mathrm{CeO}_{2}$ was mounted. These findings suggest that $\mathrm{CeO}_{2}$ is a promising candidate for future anti-icing coatings.
\end{abstract}

\section{Introduction}\label{sec:intro}
Icing on surfaces is a huge problem in society \cite{AzimiYancheshme2020, He2021, Li2023}. The ice can cause power outages and dangerous situations within transportation and communications, and removing the ice without damaging the surfaces on which it has formed can be challenging and costly \cite{AzimiYancheshme2020, He2021, Li2023, Azimi2013, Fillion2014, Latthe2019, Lian2022, Zhou2022}.  
A possible solution is to cover the ice-exposed surfaces with anti-icing coatings, and several different coatings already exist \cite{AzimiYancheshme2020, He2021, Li2023, Latthe2019, Cao2009, Kreder2016, Veronesi2021}.
It is important that these coatings are suitable matches for the systems to which they will be applied and the conditions in which they will operate \cite{Li2023, Sharifahmadian2023}. 
For example, coatings that will be used on photovoltaics need an electronic band gap allowing transmission of light within the active energy region (range of suitable excitation energies) of the photovoltaic \cite{Fillion2014}. 

A variety of techniques have been used to evaluate the anti-icing properties of coatings and their suitability for specific applications \cite{Liang2013, Wang2015, Fu2016, Mamedov2023}. However, X-ray photoelectron spectroscopy (XPS), a technique well suited to determine the chemical composition and chemical environment of the elements of surfaces receives much less attention and is often only used for checking the quality/chemical composition of the coating in question \cite{Huefner2003, Wang2013, Hofmann2016, Qi2020, Major2020, Zhang2022a, Zhang2023} or investigating the resulting surface chemistry changes post treatment \cite{Wang2013}. The ability to study water and ice with XPS is well documented and relatively straight forward \cite{Henderson2002, Fransson2016}, however, few studies are using XPS to monitor the in-situ ice formation on the coatings.

Thin ($50-400\,$nm thick) films of $\mathrm{CeO}_{2}$ can potentially be used as anti-icing coatings for photovoltaics because of their suitable band gap, abundance, and robustness \cite{Mamedov2023}. They are hydrophobic and anti-fouling, and their hydrophobicity can be enhanced by structuring the surface and/or applying hydrocarbons \cite{Mamedov2023, Khan2015}. 
Although hydrophobic coatings made from composites of $\mathrm{CeO}_{2}$ have previously been shown to be anti-icing, these studies did not include sputtered thin film $\mathrm{CeO}_{2}$ coatings and only investigated the anti-icing properties down to temperatures of $T=170\,$K \cite{Wang2015, Fu2016, Henderson2003}. 

Here we investigate the anti-icing capabilities and chemical composition of thin film $\mathrm{CeO}_{2}$ coatings simultaneously and in-situ using XPS. Measurements are performed over a wide temperature range before and after exposing various $\mathrm{CeO}_{2}$ samples to water at temperatures as low as $T\approx100\,\mathrm{K}$. In the case that no ice is observed following this preparation, the material under investigation clearly has potential for anti-icing applications, while also demonstrating the proof of concept for future XPS studies of this type.

\section{Results \& Discussion}

\begin{figure}
\begin{center}
\includegraphics[width=0.5\textwidth]{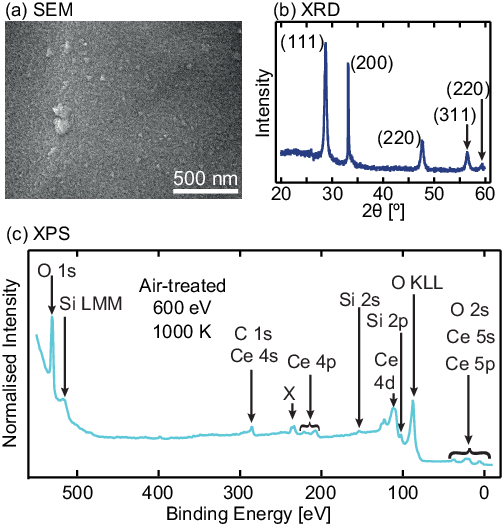}
\caption{Characterisation of the $\mathrm{CeO}_{2}$ films. (a) Scanning electron microscopy (SEM) and (b) X-ray diffraction (XRD) of a $200\,$nm thick $\mathrm{CeO}_{2}$ coating. In (b) (111), (200), (220), (331) and (222) indicate the crystallographic planes corresponding to each peak in the XRD pattern. (c) Overview X-ray photoelectron spectroscopy (XPS) measurement of a $50\,$nm $\mathrm{CeO}_{2}$ coating which has been heated at $T=870\,\mathrm{K}$ in air after growth (air-treated sample), and then to $T=1000\,\mathrm{K}$ in vacuum. The measurement was performed at $T\approx320\,\mathrm{K}$ with a photoexcitation energy of $h\nu=600\,$eV.}
\label{fig:SEM_XRD_WideXPS}
\end{center}
\end{figure}

The $\mathrm{CeO}_{2}$ coatings were grown by sputter-deposition of Ce onto a Si substrate in an oxygen-rich atmosphere (24\%) as detailed in previous works \cite{Mamedov2023}. 
Scanning electron microscopy (SEM)  and X-ray diffraction (XRD) measurements of the as-grown material are shown in Figure~\ref{fig:SEM_XRD_WideXPS} panels (a) and (b), respectively. SEM reveals that the samples are relatively uniform, whilst the XRD confirms a fcc crystal structure (space group 225) with a lattice parameter of $a=5.4\,$\AA~\cite{Gerward2005}.  
The diffraction peaks centred at $2\theta = 28.7^{\circ}$, 33.3$^{\circ}$, 47.7$^{\circ}$, 56.5$^{\circ}$ and 59.4$^{\circ}$ were identified as reflections from the (111), (200), (220), (311) and (222) planes, respectively, according to the JCPDS card \#34-039. 
The peak labelled (200) also contains a contribution from the underlying Si substrate, indicating that the Si substrate is distorted, seemingly due to internal stress caused by interactions between the substrate and the $\mathrm{CeO}_{2}$ coating \cite{Zhao2005}. 
 
XPS measurements were performed on two samples ($t\approx 50\,\mathrm{nm}$): an as-grown sample and an air-treated sample heated to $T=870\,\mathrm{K}$ in air after growth, in order to ascertain whether post heating the samples in air plays any role in how water initially is absorbed on the surface. 
Heating in an oxygen-deficient atmosphere has been shown to change the ratio of $\mathrm{Ce}^{4+}$ and $\mathrm{Ce}^{3+}$ in the coating due to a loss of O from the surface \cite{Khan2015}. 
Both samples were mounted on the same sample holder to ensure equal treatment during the experiment. 
Previous studies have shown that hydrocarbons on the surface of $\mathrm{CeO}_{2}$ can possibly cause or enhance its hydrophobicity \cite{Mamedov2023, Preston2014}. 
For samples exposed to the atmosphere, a layer of hydrocarbons and other carbon species (a.k.a. adventitious or surface carbon) normally covers the surface \cite{Barr1995}. 
In order to investigate the properties of $\mathrm{CeO}_{2}$ itself, the samples were heated to $T=1000\,\mathrm{K}$ in vacuum to remove the surface carbon. 
Figure~\ref{fig:SEM_XRD_WideXPS}(c) shows the overview XPS measurement of the air-treated sample after the in-vacuum heating. 
As expected, the most intense features in the spectra are related to the photoemission of Ce and O, together with weak signatures of Si from the underlying substrate. The peak labelled X is probably from Pr or Ta impurities in the sputtering target used to grow the $\mathrm{CeO}_{2}$, but can also be Mo from the sample holder. A small C~1s peak indicates that not all carbon species were removed during heating. 
The relative composition of Ce, O, and C atoms on the surface (Ce:O:C) was calculated from the overview XPS measurements for both samples and found to be: 1:2:4 before and 1:2:2 after in-vacuum heating for the as-grown sample and 1:3:8 before and 1:2:1 after in-vacuum heating for the air-treated sample. 
For both samples, there are 2-3 O-atoms for every Ce-atom, which is as expected for $\mathrm{CeO}_{2}$ grown by sputter deposition \cite{Khan2015}. A small reduction in the photoemission intensity from O~1s is observed consistent with the change from $\mathrm{Ce}^{4+}$ to $\mathrm{Ce}^{3+}$ as previously discussed. 
A larger reduction of carbon species is observed on the air-treated sample relative to the as-grown sample following heating in vacuum, which could indicate that surface carbon is more loosely bound. For both samples, however, the surface carbon is significantly reduced.

Since the aim was to investigate the early stages of water exposure, the samples were cooled to $T=100\,\mathrm{K}$, and $\mathrm{H}_{2}\mathrm{O}$ was leaked into the vacuum system at a constant pressure of $1\times10^{-8}\,\mathrm{mbar}$ for 10\,min. This is equivalent to a dose of $(5\pm1)\,$ Langmuir (atomic layers) \cite{Lueth2013}. 
Assuming a sticking coefficient of $1$, i.e. that all the deposited $\mathrm{H}_{2}\mathrm{O}$ adheres to the cold surface, this corresponds to a $(1.4\pm0.3)\,\mathrm{nm}$ thick layer of ice. 

\begin{figure*}
\begin{center}
\includegraphics[width=\textwidth]{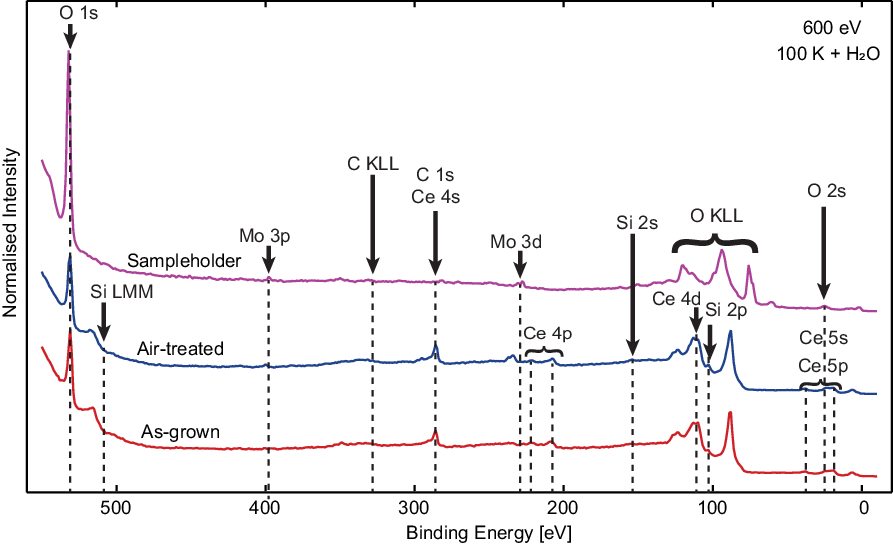}
\caption{Overview X-ray photoelectron spectroscopy (XPS) measurement of two $50\,$nm thick $\mathrm{CeO}_{2}$ coatings (as-grown and air-treated), and their Mo sample holder, after exposure to $\mathrm{H}_{2}\mathrm{O}$. The samples and the sample holder were kept at $T\approx100\,\mathrm{K}$ during the exposure and the measurement, and measured with a photoexcitation energy of $h\nu=600\,$eV.}
\label{fig:WaterDepWide}
\end{center}
\end{figure*} 

Overview spectra of both samples measured after exposure to $\mathrm{H}_{2}\mathrm{O}$ are shown in Figure~\ref{fig:WaterDepWide}. It can be seen that there is little difference between the air-treated and as-grown samples, or any notable change from the air-treated spectra presented in Figure~\ref{fig:SEM_XRD_WideXPS}(c). Hence, there appears to be no ice (or very little ice) on the surface of either of the $\mathrm{CeO}_{2}$ samples. A rough quantification of the overview spectra in Figure~\ref{fig:WaterDepWide} yields Ce:O ratios of 1:4 (as-grown) and 1:2 (air-treated). 
In other words, there is no significant increase in O on the sample surfaces after exposure to $\mathrm{H}_{2}\mathrm{O}$  compared to just after heating. It can therefore be inferred that the O~1s signal is mainly due to O in $\mathrm{CeO}_{2}$, and not from ice on the $\mathrm{CeO}_{2}$ surface. 
Contrarily, the Mo sample holder has a very large photoemission signal originating from O. The O~1s and O~KLL components are too large relative to the Mo peaks to allow them to be attributed to Mo oxides.  
Furthermore, the binding energy of the Mo~3d peak agrees with the accepted binding energy of metallic Mo but is lower than that of oxidised Mo, suggesting that the Mo sample holder is not fully oxidised \cite{Brox1988}. Consequently, the vast majority of the intensity from O must be attributed to originate from the ice layer.  
The thickness of the ice layer based on the photoemission intensity is calculated as $\approx 2.7\,\mathrm{nm}$ (see the Supporting Information \cite{SupplMat} for details).
These measurements suggest that the ice layer is thicker than our earlier prediction of $\approx 1.4\,\mathrm{nm}$ estimated from the dose of $\mathrm{H}_{2}\mathrm{O}$.  
However, the two estimates are consistent within a factor of 2. Given the assumptions of these simple estimates (for example, both assume a uniform overlayer thickness, which is likely not the case in reality), one can expect a significant uncertainty (see further details in the Supporting Information \cite{SupplMat}). Thus, the thickness estimates give a useful order-of-magnitude estimate of the ice thickness, but should not be taken as precise values. Ultimately, it is safe to conclude that the XPS measurements clearly show that there is ice on the sample holder, but not on the $\mathrm{CeO}_{2}$ surfaces. 

\begin{figure}
\begin{center}
\includegraphics[width=0.5\textwidth]{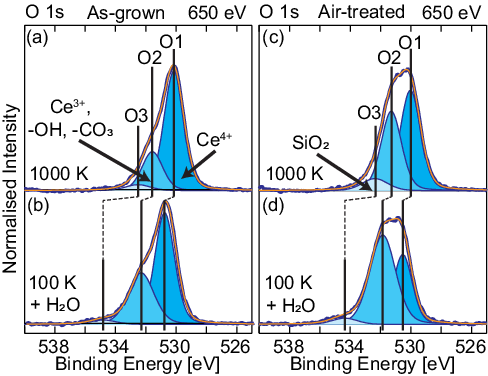}
\caption{Photoemission intensity of the O~1s core level of the as-grown ((a) and (b)) and air-treated ((c) and (d)) samples measured with photoexcitation energy $h\nu=650\,$eV. (a) and (c) were measured at $T\approx320\,$K after heating to $T=1000\,$K in vacuum. (b) and (d) were measured at $T\approx100\,$K after $\mathrm{H}_{2}\mathrm{O}$ exposure. The O1 peak (dark blue) represents $\mathrm{Ce}^{4+}$, the O2 peak (medium blue) represents $\mathrm{Ce}^{3+}$, $\mathrm{-OH}$ and $\mathrm{-CO}_{3}$, and the O3 peak (light blue) represents $\mathrm{SiO}_2$.}
\label{fig:O1sSmall}
\end{center}
\end{figure}

\begin{figure}
\begin{center}
\includegraphics[width=0.5\textwidth]{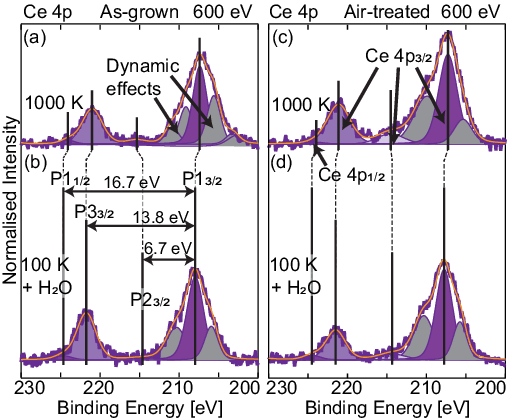}
\caption{Photoemission intensity of the Ce~4p core level of the as-grown ((a) and (b)) and air-treated ((c) and (d)) samples measured with photoexcitation energy $h\nu=600\,$eV. (a) and (c) were measured at $T\approx320\,$K after heating in to $T=1000\,$K in vacuum. (b) and (d) were measured at $T\approx100\,$K after $\mathrm{H}_{2}\mathrm{O}$ exposure. The measured envelope is deconvoluted into one spin-orbit coupled doublet, $\mathrm{P1}_{3/2}$ and $\mathrm{P1}_{1/2}$ (dark purple), a charge transfer related satellite, $\mathrm{P2}_{3/2}$ (light purple), and a shake-up satellite, $\mathrm{P3}_{3/2}$ (medium purple). The peaks labelled dynamic effects (grey) are assigned to Coster-Kronig transitions. }
\label{fig:Ce4pSmall}
\end{center}
\end{figure}

High-resolution spectra of the O~1s and Ce~4p core-level from throughout the experiments are presented in Figures~\ref{fig:O1sSmall} and \ref{fig:Ce4pSmall}, respectively. Deconvolution of the photoemission profiles is undertaken through fitting pseudo-Voigt line shapes and Shirley background step functions. The intensity is then normalised to the total background intensity \cite{SupplMat}.  As is shown in Figure~\ref{fig:O1sSmall} the measured profile required 3 components to fit and are assigned as follows: O1 (dark blue) corresponding to photoemission from $\mathrm{Ce}^{4+}$, O2 (medium blue) is assigned to photoemission from  $\mathrm{Ce}^{3+}$ along with hydroxyls ($\mathrm{-OH}$) and carbonates ($\mathrm{-CO}_{3}$) similar to References \cite{Khan2015}, \cite{Preisler2001}, \cite{Martinez2011} and \cite{Maslakov2018}. 
The component labelled O3 (light blue) is ascribed to $\mathrm{SiO}_2$ from the underlying oxidised Si substrate \cite{Khan2015, Preisler2001, Martinez2011}. 

As discussed earlier, the reduction of $\mathrm{Ce}^{4+}$ to $\mathrm{Ce}^{3+}$ during air treatment is known to occur, and is further confirmed by the higher intensity of the component relating to $\mathrm{Ce}^{3+}$ in the O~1s profile for the air-treated sample in Figure~\ref{fig:O1sSmall} panels (c) and (d), compared to the as-grown sample shown in panels (a) and (b). This reduction is also known to occur for prolonged exposure to ultra-high vacuum (UHV) environments \cite{Khan2015,Zhang2004} hence, an increase in the $\mathrm{Ce}^{3+}$ component is expected during the experiment. 
Although the increase in  $\mathrm{Ce}^{3+}$ is slightly higher in our measurements than that found by Khan \textit{et al.} \cite{Khan2015}, $\mathrm{H}_{2}\mathrm{O}$ exposure is not expected to result in further reduction at temperatures below $650\,$K \cite{Henderson2003}. The intensity increase could also be due to an increased amount of $\mathrm{-CO}_{3}$ or $\mathrm{-OH}$, however, from the C~1s core level (see Figure~S1 in the Supporting Information \cite{SupplMat}) the amount of $\mathrm{-CO}_{3}$ was found to decrease, and the amount of $\mathrm{C-OH}$ increased only slightly. Finally, the component relating to the oxidised Si substrate is reduced or unchanged after $\mathrm{H}_{2}\mathrm{O}$ exposure. Considering all of these factors it is difficult to assign any of the observed component variations to the growth of an ice layer on either of the $\mathrm{CeO}_{2}$ samples.

The photoemission intensity from the Ce~4p region shown in Figure~\ref{fig:Ce4pSmall} is similar to previous measurements \cite{Mamedov2023, Maslakov2018, Burroughs1976}, and changes very little after $\mathrm{H}_{2}\mathrm{O}$ exposure. 
$\mathrm{CeO}_{2}$ has mixed valency, meaning that its 4f orbital is close to the Fermi-level and can either be occupied or unoccupied \cite{Fujimori1983a, Fujimori1983}. As a consequence of this, there are atoms with different electronic configurations both in the ground state and after the creation of a photohole during the photoemission process \cite{Fujimori1983a, Fujimori1983}. The measured kinetic energy of the emitted electron will depend on the final state electronic configuration \cite{Fujimori1983}. Thus, the mixed valency causes additional peaks and complex intensity variations in the measured spectra \cite{Maslakov2018, Fujimori1983a, Fujimori1983}.  
Even if such final state effects are not commonly seen in XPS, they are well known for $\mathrm{CeO}_{2}$ and similar oxides of f-block elements \cite{Mamedov2023, Maslakov2018, Fujimori1983a, Fujimori1983, Kotani1987, Monsen2012}. 
For $\mathrm{CeO}_{2}$ the different final states result in three main doublets, P1, P2 and P3, shown in Figure~\ref{fig:Ce4pSmall} as dark ($\mathrm{P1}_{3/2}$ and $\mathrm{P1}_{1/2}$), light ($\mathrm{P2}_{3/2}$) and medium ($\mathrm{P3}_{3/2}$) purple, respectively  \cite{Mamedov2023, Maslakov2018, Fujimori1983}. The separations between the peaks are reported in Table~\ref{tab:Separation} and predominately agree with those found in literature \cite{Maslakov2018}. The separations labelled with an asterisk are larger than previously reported, however, the uncertainty in their position is larger at this stage in the experiment, as the intensity of the $\mathrm{P2}_{3/2}$ peak is small and difficult to distinguish from the other peaks related to dynamic effects.

The peaks labelled ``dynamic effects'' (grey) are thought to be caused by Coster-Kronig transitions for which the final state photoholes are in the same main shell as the emitted electron \cite{Maslakov2018, Nyholm1981}. The total intensity of peaks related to Coster-Kronig transitions is somewhat lower at $100\,$K compared to at $1000\,$K, suggesting that there are fewer Coster-Kronig transitions after cooling and $\mathrm{H}_{2}\mathrm{O}$ exposure.

\begin{table}
\caption{Energy separation between the peaks in the Ce~4p spectra.} \label{tab:Separation}
\begin{indented}
\lineup
\item[]\begin{tabular}{@{}lllll}	
\br	
Separation & \multicolumn{2}{c}{As-grown} & \multicolumn{2}{c}{Air-treated} \\ 
  between & $1000\,$K & $100\,$K$+\mathrm{H}_{2}\mathrm{O}$ & $1000\,$K & $100\,$K$+\mathrm{H}_{2}\mathrm{O}$ \\
\mr
$\mathrm{P1}_{3/2}$ and $\mathrm{P1}_{1/2}$ & $16.7\,$eV	&	$16.7\,$eV  &	 $16.7\,$eV	&	$16.7\,$eV\\
$\mathrm{P1}_{3/2}$ and $\mathrm{P2}_{3/2}$	&  $\08.0\,$eV\textsuperscript{*}	&	$\06.7\,$eV&	 $\07.3\,$eV\textsuperscript{*}	&	$\06.5\,$eV\\
$\mathrm{P1}_{3/2}$ and $\mathrm{P3}_{3/2}$ & $13.6\,$eV	&	$13.8\,$eV &	 $13.8\,$eV	&	$13.7\,$eV\\	 
\br	
\end{tabular}
\end{indented}
\end{table}

XPS measurements showing the evolution of the valence band maximum and O~1s, Ce~4p, and C~1s core levels after the samples have been loaded into the vacuum chamber, heated to increasingly higher temperatures, and cooled and exposed to $\mathrm{H}_{2}\mathrm{O}$ have been included in the Supporting Information \cite{SupplMat}. 
These spectra shift towards lower binding energies when heated, and shift (partly) back upon cooling and $\mathrm{H}_{2}\mathrm{O}$ exposure.
Binding energy shifts have also been observed in previous studies and were explained by the transition from $\mathrm{Ce}^{4+}$ to $\mathrm{Ce}^{3+}$ \cite{Preisler2001, Maslakov2018}. However, there the shifts were towards higher binding energies upon heating in vacuum, i.e. opposite to the shifts observed in the spectra presented in this manuscript \cite{Preisler2001, Maslakov2018}. 
Possible explanations of the shifts include (temperature dependent) changes in the electrostatic potential at the surface or at the interface between the $\mathrm{CeO}_{2}$ coating and the Si substrate, potentially caused by adsorbates, band bending, or resistivity changes of the $\mathrm{CeO}_{2}$ \cite{SupplMat, Kampen:2003, Cabailh:2004}. 
The spectra from the as-grown sample shifted only partly back to higher binding energies when it was cooled, indicating that heating removed some surface adsorbates which acted as $n$-type dopants before the surface was heated.

\section{Conclusion}
$\mathrm{CeO}_{2}$ coatings have been investigated before and after exposing the cold sample surfaces to $(5\pm1)\,$ Langmuir of $\mathrm{H}_{2}\mathrm{O}$, equivalent to $(1.4\pm0.3)\,\mathrm{nm}$ of ice.  
Via analysis of high-resolution core level spectroscopy, it was shown that ice does not form on the sample surfaces, even if the measurements indicated that a $\approx 2.7\,\mathrm{nm}$ thick layer of ice had formed on the surrounding areas.   
Both the O~1s and Ce~4p spectra have their main component related to $\mathrm{CeO}_{2}$ and show little change before and after exposure to $\mathrm{H}_{2}\mathrm{O}$. The film that received a post-growth in-air treatment appears to have a slightly higher proportion of $\mathrm{Ce}^{3+}$ compared to the as-grown film, however, this does not seem to have a noticeable effect on the anti-icing properties. 
Since post-growth treatment is not required, the coatings can be made more efficiently and affordable compared to many other coatings \cite{Li2023, Cao2009} making them good candidates for anti-icing applications. 
Additionally, it has been demonstrated that XPS is a sensitive and suitable method to evaluate the early stages of ice-formation or anti-icing behaviour of surfaces and coatings.

\section{Acknowledgements}
This work was partly supported by the Research Council of Norway, project numbers 262~633, 309~827, and 335~022. 
The authors acknowledge the staff at the ASTRID2 synchrotron in Aarhus, Denmark, for providing access to their synchrotron radiation facilities, and for practical assistance and discussions.

\section*{References}

\bibliographystyle{unsrt}
\bibliography{CeO2TempDep} 

\begin{thebibliography}{10}

\bibitem{AzimiYancheshme2020}
Amir Azimi~Yancheshme, Anahita Allahdini, Khosrow Maghsoudi, Reza Jafari, and
  Gelareh Momen.
\newblock Potential anti-icing applications of encapsulated phase change
  material–embedded coatings; a review.
\newblock {\em J. Energy Storage}, 31:101638, October 2020.

\bibitem{He2021}
Hua He and Zhiguang Guo.
\newblock Superhydrophobic materials used for anti-icing {Theory}, application,
  and development.
\newblock {\em iScience}, 24(11):103357, November 2021.

\bibitem{Li2023}
Bo~Li, Jie Bai, Jinhang He, Chao Ding, Xu~Dai, Wenjun Ci, Tao Zhu, Ruijin Liao,
  and Yuan Yuan.
\newblock A {Review} on {Superhydrophobic} {Surface} with {Anti}-{Icing}
  {Properties} in {Overhead} {Transmission} {Lines}.
\newblock {\em Coatings}, 13(2):301, February 2023.

\bibitem{Azimi2013}
Gisele Azimi, Rajeev Dhiman, Hyuk-Min Kwon, Adam~T. Paxson, and Kripa~K.
  Varanasi.
\newblock Hydrophobicity of rare-earth oxide ceramics.
\newblock {\em Nat. Mater.}, 12(4):315--320, April 2013.

\bibitem{Fillion2014}
R.~M. Fillion, A.~R. Riahi, and A.~Edrisy.
\newblock A review of icing prevention in photovoltaic devices by surface
  engineering.
\newblock {\em Renewable Sustainable Energy Rev.}, 32:797--809, April 2014.

\bibitem{Latthe2019}
Sanjay~S. Latthe, Rajaram~S. Sutar, Appasaheb~K. Bhosale, Saravanan Nagappan,
  Chang-Sik Ha, Kishor~Kumar Sadasivuni, Shanhu Liu, and Ruimin Xing.
\newblock Recent developments in air-trapped superhydrophobic and
  liquid-infused slippery surfaces for anti-icing application.
\newblock {\em Prog. Org. Coat.}, 137:105373, December 2019.

\bibitem{Lian2022}
Chengxing Lian, Christopher Emersic, Fatema~H. Rajab, Ian Cotton, Xu~Zhang,
  Robert Lowndes, and Lin Li.
\newblock Assessing the {Superhydrophobic} {Performance} of {Laser}
  {Micropatterned} {Aluminium} {Overhead} {Line} {Conductor} {Material}.
\newblock {\em IEEE Trans. Power Delivery}, 37(2):972--979, April 2022.

\bibitem{Zhou2022}
Liang Zhou, Ruidi Liu, and Xian Yi.
\newblock Research and development of anti-icing/deicing techniques for
  vessels: Review.
\newblock {\em Ocean Eng.}, 260:112008, 2022.

\bibitem{Cao2009}
Liangliang Cao, Andrew~K. Jones, Vinod~K. Sikka, Jianzhong Wu, and Di~Gao.
\newblock Anti-{Icing} {Superhydrophobic} {Coatings}.
\newblock {\em Langmuir}, 25(21):12444--12448, November 2009.

\bibitem{Kreder2016}
Michael~J. Kreder, Jack Alvarenga, Philseok Kim, and Joanna Aizenberg.
\newblock Design of anti-icing surfaces: smooth, textured or slippery?
\newblock {\em Nat. Rev. Mater.}, 1(1):1--15, January 2016.

\bibitem{Veronesi2021}
Federico Veronesi, Giulio Boveri, Julio Mora, Alessandro Corozzi, and Mariarosa
  Raimondo.
\newblock Icephobic properties of anti-wetting coatings for aeronautical
  applications.
\newblock {\em Surf. Coat. Technol.}, 421:127363, sep 2021.

\bibitem{Sharifahmadian2023}
Omid Sharifahmadian, Amirhossein Pakseresht, Saeed Mirzaei, Marek Eliáš, and
  Dušan Galusek.
\newblock Mechanically robust hydrophobic fluorine-doped diamond-like carbon
  film on glass substrate.
\newblock {\em Diamond Relat. Mater.}, 138:110252, October 2023.

\bibitem{Liang2013}
Jin Liang, Yunchu Hu, Youhua Fan, and Hong Chen.
\newblock Formation of superhydrophobic cerium oxide surfaces on aluminum
  substrate and its corrosion resistance properties.
\newblock {\em Surf. Interface Anal.}, 45(8):1211--1216, 2013.

\bibitem{Wang2015}
Pengren Wang, Chaoyi Peng, Binrui Wu, Zhiqing Yuan, Fubiao Yang, and Jingcheng
  Zeng.
\newblock A facile method of fabricating mechanical durable anti-icing coatings
  based on {CeO}$_2$ microparticles.
\newblock {\em IOP Conf. Ser.: Mater. Sci. Eng.}, 87(1):012062, June 2015.

\bibitem{Fu2016}
Sin-Pui Fu, Rakesh~P. Sahu, Estefan Diaz, Jaqueline~Rojas Robles, Chen Chen,
  Xue Rui, Robert~F. Klie, Alexander~L. Yarin, and Jeremiah~T. Abiade.
\newblock Dynamic study of liquid drop impact on supercooled cerium dioxide:
  Anti-icing behavior.
\newblock {\em Langmuir}, 32(24):6148--6162, may 2016.

\bibitem{Mamedov2023}
D.~Mamedov, A.~C. Åsland, S.~P. Cooil, H.~I. Røst, J.~Bakkelund,
  A.~Allaniyazov, J.~W. Wells, and S.~Karazhanov.
\newblock Enhanced hydrophobicity of {CeO}$_2$ thin films: {Role} of the
  morphology, adsorbed species and crystallography.
\newblock {\em Mater. Today Commun.}, 35:106323, June 2023.

\bibitem{Huefner2003}
Stefan Hüfner.
\newblock {\em Photoelectron Spectroscopy}.
\newblock Springer Berlin Heidelberg, 2003.

\bibitem{Wang2013}
Yuanyi Wang, Jian Xue, Qingjun Wang, Qingmin Chen, and Jianfu Ding.
\newblock Verification of {Icephobic}/{Anti}-icing {Properties} of a
  {Superhydrophobic} {Surface}.
\newblock {\em ACS Appl. Mater. Interfaces}, 5(8):3370--3381, April 2013.

\bibitem{Hofmann2016}
{Ph. Hofmann}.
\newblock {\em Surface {Physics}: {An} {Introduction}}.
\newblock Philip Hofmann, 2016.

\bibitem{Qi2020}
Yanli Qi, Zhangbin Yang, Tingting Chen, Yulin Xi, and Jun Zhang.
\newblock Fabrication of superhydrophobic surface with desirable anti-icing
  performance based on micro/nano-structures and organosilane groups.
\newblock {\em Appl. Surf. Sci.}, 501:144165, January 2020.

\bibitem{Major2020}
George~H. Major, Neal Fairley, Peter M.~A. Sherwood, Matthew~R. Linford, Jeff
  Terry, Vincent Fernandez, and Kateryna Artyushkova.
\newblock Practical guide for curve fitting in x-ray photoelectron
  spectroscopy.
\newblock {\em J. Vac. Sci. Technol., A}, 38(6):061203, December 2020.

\bibitem{Zhang2022a}
Binbin Zhang, Mengying Qiao, Weichen Xu, and Baorong Hou.
\newblock All-organic superhydrophobic coating with anti-corrosion, anti-icing
  capabilities and prospective marine atmospheric salt-deliquesce self-coalesce
  protective mechanism.
\newblock {\em J. Ind. Eng. Chem.}, 115:430--439, nov 2022.

\bibitem{Zhang2023}
Lin-Bo Zhang, Han-Xuan Zhang, Zhi-Jie Liu, Xian-Yu Jiang, Simeon Agathopoulos,
  Zhou Deng, Hao-Yu Gao, Li~Zhang, Hai-Peng Lu, Long-Jiang Deng, and Liang-Jun
  Yin.
\newblock Nano-silica anti-icing coatings for protecting wind-power turbine fan
  blades.
\newblock {\em J. Colloid Interface Sci.}, 630:1--10, January 2023.

\bibitem{Henderson2002}
Michael~A. Henderson.
\newblock The interaction of water with solid surfaces: fundamental aspects
  revisited.
\newblock {\em Surf. Sci. Rep.}, 46(1):1--308, May 2002.

\bibitem{Fransson2016}
Thomas Fransson, Yoshihisa Harada, Nobuhiro Kosugi, Nicholas~A. Besley, Bernd
  Winter, John~J. Rehr, Lars G.~M. Pettersson, and Anders Nilsson.
\newblock X-ray and {Electron} {Spectroscopy} of {Water}.
\newblock {\em Chem. Rev.}, 116(13):7551--7569, jul 2016.

\bibitem{Khan2015}
Sami Khan, Gisele Azimi, Bilge Yildiz, and Kripa~K. Varanasi.
\newblock Role of surface oxygen-to-metal ratio on the wettability of
  rare-earth oxides.
\newblock {\em Appl. Phys. Lett.}, 106(6):061601, February 2015.

\bibitem{Henderson2003}
M.~A. Henderson, C.~L. Perkins, M.~H. Engelhard, S.~Thevuthasan, and C.~H.~F.
  Peden.
\newblock Redox properties of water on the oxidized and reduced surfaces of
  {CeO}$_2$(111).
\newblock {\em Surf. Sci.}, 526(1):1--18, February 2003.

\bibitem{Gerward2005}
L.~Gerward, J.~Staun~Olsen, L.~Petit, G.~Vaitheeswaran, V.~Kanchana, and
  A.~Svane.
\newblock Bulk modulus of {CeO}$_2$ and {PrO}$_2$—{An} experimental and
  theoretical study.
\newblock {\em J. Alloys Compd.}, 400(1):56--61, September 2005.

\bibitem{Zhao2005}
Lili Zhao, Martin Steinhart, Maekele Yosef, Sung~Kyun Lee, Torsten Geppert,
  Eckhard Pippel, Roland Scholz, Ulrich Gösele, and Sabine Schlecht.
\newblock Lithium {Niobate} {Microtubes} within {Ordered} {Macroporous}
  {Silicon} by {Templated} {Thermolysis} of a {Single} {Source} {Precursor}.
\newblock {\em Chem. Mater.}, 17(1):3--5, January 2005.

\bibitem{Preston2014}
Daniel~J. Preston, Nenad Miljkovic, Jean Sack, Ryan Enright, John Queeney, and
  Evelyn~N. Wang.
\newblock Effect of hydrocarbon adsorption on the wettability of rare earth
  oxide ceramics.
\newblock {\em Appl. Phys. Lett.}, 105(1), jul 2014.

\bibitem{Barr1995}
Tery~L. Barr and Sudipta Seal.
\newblock Nature of the use of adventitious carbon as a binding energy
  standard.
\newblock {\em J. Vac. Sci. Technol., A}, 13(3):1239--1246, may 1995.

\bibitem{Lueth2013}
Hans Lüth.
\newblock {\em Surfaces and {Interfaces} of {Solid} {Materials}}.
\newblock Springer Science \& Business Media, March 2013.

\bibitem{Brox1988}
B.~Brox and I.~Olefjord.
\newblock {ESCA} {Studies} of {MoO}$_2$ and {MoO}$_3$.
\newblock {\em Surf. Interface Anal.}, 13(1):3--6, 1988.

\bibitem{SupplMat}
See the supporting information at [insert link] including references
  \cite{Mamedov2023, Huefner2003, Khan2015, Gerward2005, Preston2014,
  Preisler2001, Martinez2011, Maslakov2018, Burroughs1976, Zangwill1988,
  Sarapatka1992, Sassaroli2004, Song2012, AcostaSilva2017, Zemlyanov2018,
  Powell2020, Astley2022, Xu2022} and information about the the experimental
  details, further details about the normalisation and the ice thickness
  calculation from the xps, additional spectra measured after annealing the
  samples to increasingly higher temperatures, and a discussion of the binding
  energy shifts of the spectra.

\bibitem{Preisler2001}
E.~J. Preisler, O.~J. Marsh, R.~A. Beach, and T.~C. McGill.
\newblock Stability of cerium oxide on silicon studied by x-ray photoelectron
  spectroscopy.
\newblock {\em J. Vac. Sci. Technol., B: Microelectron. Nanometer Struct.
  Process., Meas., Phenom.}, 19(4):1611--1618, jul 2001.

\bibitem{Martinez2011}
L.~Martínez, E.~Román, J.~L. de~Segovia, S.~Poupard, J.~Creus, and
  F.~Pedraza.
\newblock Surface study of cerium oxide based coatings obtained by cathodic
  electrodeposition on zinc.
\newblock {\em Appl. Surf. Sci.}, 257(14):6202--6207, May 2011.

\bibitem{Maslakov2018}
Konstantin~I. Maslakov, Yury~A. Teterin, Aleksej~J. Popel, Anton~Yu. Teterin,
  Kirill~E. Ivanov, Stepan~N. Kalmykov, Vladimir~G. Petrov, Peter~K. Petrov,
  and Ian Farnan.
\newblock {XPS} study of ion irradiated and unirradiated {CeO}$_2$ bulk and
  thin film samples.
\newblock {\em Appl. Surf. Sci.}, 448:154--162, August 2018.

\bibitem{Zhang2004}
Feng Zhang, Peng Wang, J~Koberstein, S~Khalid, and Siu-Wai Chan.
\newblock Cerium oxidation state in ceria nanoparticles studied with {X}-ray
  photoelectron spectroscopy and absorption near edge spectroscopy.
\newblock {\em Surf. Sci.}, 563(1):74--82, August 2004.

\bibitem{Burroughs1976}
Peter Burroughs, Andrew Hamnett, Anthony~F. Orchard, and Geoffrey Thornton.
\newblock Satellite structure in the {X}-ray photoelectron spectra of some
  binary and mixed oxides of lanthanum and cerium.
\newblock {\em J. Chem. Soc., Dalton Trans.}, (17):1686--1698, jan 1976.

\bibitem{Fujimori1983a}
Atsushi Fujimori.
\newblock 4f- and core-level photoemission satellites in cerium compounds.
\newblock {\em Phys. Rev. B}, 27(7):3992--4001, apr 1983.

\bibitem{Fujimori1983}
Atsushi Fujimori.
\newblock Mixed-valent ground state of $\mathrm{Ce}\mathrm{O}_{2}$.
\newblock {\em Phys. Rev. B}, 28(4):2281--2283, August 1983.

\bibitem{Kotani1987}
A.~Kotani, M.~Okada, and T.~Jo.
\newblock Many {Body} {Effects} in {Core} {Level} {Spectra} of {Lanthanum}
  {Compounds}.
\newblock {\em Phys. Scr.}, 35(4):566--569, April 1987.

\bibitem{Monsen2012}
{\AA}.F. Monsen, F.~Song, Z.S. Li, J.E. Boschker, T.~Tybell, E.~Wahlstr{\oe}m,
  and J.W. Wells.
\newblock Surface stoichiometry of $la_{0.7}sr_{0.3}mno_{3}$ during in vacuo
  preparation: A synchrotron photoemission study.
\newblock {\em Surf. Sci.}, 606(17-18):1360--1366, sep 2012.

\bibitem{Nyholm1981}
R.~Nyholm, N.~Martensson, A.~Lebugle, and U.~Axelsson.
\newblock Auger and {Coster}-{Kronig} broadening effects in the 2p and 3p
  photoelectron spectra from the metals $^{22}\mathrm{Ti}-^{30}\mathrm{Zn}$.
\newblock {\em J. Phys. F: Met. Phys.}, 11(8):1727--1733, August 1981.

\bibitem{Kampen:2003}
T.~U. Kampen, G.~Gavrila, H.~Mendez, D.~R.~T. Zahn, A.~R. Vearey-Roberts, D.~A.
  Evans, J.~Wells, I.~McGovern, and W.~Braun.
\newblock Electronic properties of interfaces between perylene derivatives and
  {GaAs}(001) surfaces.
\newblock {\em J. Phys. Cond Matt.}, 15(38):S2679--S2692, 2003.

\bibitem{Cabailh:2004}
G.~Cabailh, J.~W. Wells, I.~T. McGovern, A.~R. Vearey-Roberts, A.~Bushell, and
  D.~A. Evans.
\newblock Synchrotron radiation studies of the growth and beam damage of
  tin-phthalocyanine on {G}a{A}s(001)-1x6 substrates.
\newblock {\em Appl. Surf. Sci.}, 234(1-4):144--148, 2004.

\bibitem{Zangwill1988}
Andrew Zangwill.
\newblock {\em Physics at Surfaces}.
\newblock Cambridge University Press, mar 1988.

\bibitem{Sarapatka1992}
T.~J. Šarapatka.
\newblock {XPS} study of lead on {SiOx}/{Si}: temperature and light effects.
\newblock {\em Surf. Sci.}, 275(3):443--449, September 1992.

\bibitem{Sassaroli2004}
Angelo Sassaroli and Sergio Fantini.
\newblock Comment on the modified {Beer}–{Lambert} law for scattering media.
\newblock {\em Phys. Med. Biol.}, 49(14):N255, July 2004.

\bibitem{Song2012}
F.~Song, {\AA}.~Monsen, Z.~S. Li, E.~M. Choi, J.~L. MacManus-Driscoll,
  J.~Xiong, Q.~X. Jia, E.~Wahlstr{\oe}m, and J.~W. Wells.
\newblock Extracting the near surface stoichiometry of
  {BiFe}$0.5${Mn}$_0.5${O}$_3$ thin films; a finite element maximum entropy
  approach.
\newblock {\em Surf. Sci.}, 606(23):1771--1776, December 2012.

\bibitem{AcostaSilva2017}
Yuliana de~Jesus Acosta-Silva, Rebeca Castañedo-Perez, Gerardo Torres-Delgado,
  Arturo Méndez-López, and Orlando Zelaya-Ángel.
\newblock Effect of annealing temperature on structural, morphological and
  optical properties of {CeO}$_2$ thin films obtained from a simple precursor
  solution.
\newblock {\em J. Sol-Gel Sci. Technol.}, 82(1):20--27, April 2017.

\bibitem{Zemlyanov2018}
Dmitry~Y. Zemlyanov, Michael Jespersen, Dmitry~N. Zakharov, Jianjun Hu, Rajib
  Paul, Anurag Kumar, Shanee Pacley, Nicholas Glavin, David Saenz, Kyle~C.
  Smith, Timothy~S. Fisher, and Andrey~A. Voevodin.
\newblock Versatile technique for assessing thickness of {2D} layered materials
  by {XPS}.
\newblock {\em Nanotechnology}, 29(11), February 2018.

\bibitem{Powell2020}
Cedric~J. Powell.
\newblock Practical guide for inelastic mean free paths, effective attenuation
  lengths, mean escape depths, and information depths in x-ray photoelectron
  spectroscopy.
\newblock {\em J. Vac. Sci. Technol., A}, 38(2):023209, February 2020.

\bibitem{Astley2022}
Simon Astley, Di~Hu, Kerry Hazeldine, Johnathan Ash, Rachel E. Cross, Simon
  Cooil, Martin W. Allen, James Evans, Kelvin James, Federica Venturini, David
  C. Grinter, Pilar Ferrer, Rosa Arrigo, Georg Held, Gruffudd T. Williams,
  and D.~Andrew Evans.
\newblock Identifying chemical and physical changes in wide-gap semiconductors
  using real-time and near ambient-pressure {XPS}.
\newblock {\em Faraday Discuss.}, 236(0):191--204, 2022.

\bibitem{Xu2022}
Tao Xu, Kræn~C. Adamsen, Zheshen Li, Lutz Lammich, Jeppe~V. Lauritsen, and
  Stefan Wendt.
\newblock {WO}$_3$ {Monomers} {Supported} on {Anatase} {TiO}$_2$(101), -(001),
  and {Rutile} {TiO}$_2$(110): {A} {Comparative} {STM} and {XPS} {Study}.
\newblock {\em J. Phys. Chem. C}, 126(5):2493--2502, February 2022.

\end{thebibliography}


\begin{thebibliography}{10}

\bibitem{Gerward2005}
L.~Gerward, J.~Staun~Olsen, L.~Petit, G.~Vaitheeswaran, V.~Kanchana, and
  A.~Svane.
\newblock Bulk modulus of {CeO}$_2$ and {PrO}$_2$—{An} experimental and
  theoretical study.
\newblock {\em J. Alloys Compd.}, 400(1):56--61, September 2005.

\bibitem{Xu2022}
Tao Xu, Kræn~C. Adamsen, Zheshen Li, Lutz Lammich, Jeppe~V. Lauritsen, and
  Stefan Wendt.
\newblock {WO}$_3$ {Monomers} {Supported} on {Anatase} {TiO}$_2$(101), -(001),
  and {Rutile} {TiO}$_2$(110): {A} {Comparative} {STM} and {XPS} {Study}.
\newblock {\em J. Phys. Chem. C}, 126(5):2493--2502, February 2022.

\bibitem{Sassaroli2004}
Angelo Sassaroli and Sergio Fantini.
\newblock Comment on the modified {Beer}–{Lambert} law for scattering media.
\newblock {\em Phys. Med. Biol.}, 49(14):N255, July 2004.

\bibitem{Song2012}
F.~Song, {\AA}.~Monsen, Z.~S. Li, E.~M. Choi, J.~L. MacManus-Driscoll,
  J.~Xiong, Q.~X. Jia, E.~Wahlstr{\oe}m, and J.~W. Wells.
\newblock Extracting the near surface stoichiometry of
  {BiFe}$0.5${Mn}$_0.5${O}$_3$ thin films; a finite element maximum entropy
  approach.
\newblock {\em Surf. Sci.}, 606(23):1771--1776, December 2012.

\bibitem{Zemlyanov2018}
Dmitry~Y. Zemlyanov, Michael Jespersen, Dmitry~N. Zakharov, Jianjun Hu, Rajib
  Paul, Anurag Kumar, Shanee Pacley, Nicholas Glavin, David Saenz, Kyle~C.
  Smith, Timothy~S. Fisher, and Andrey~A. Voevodin.
\newblock Versatile technique for assessing thickness of {2D} layered materials
  by {XPS}.
\newblock {\em Nanotechnology}, 29(11), February 2018.

\bibitem{Zangwill1988}
Andrew Zangwill.
\newblock {\em Physics at Surfaces}.
\newblock Cambridge University Press, mar 1988.

\bibitem{Powell2020}
Cedric~J. Powell.
\newblock Practical guide for inelastic mean free paths, effective attenuation
  lengths, mean escape depths, and information depths in x-ray photoelectron
  spectroscopy.
\newblock {\em J. Vac. Sci. Technol., A}, 38(2):023209, February 2020.

\bibitem{Preston2014}
Daniel~J. Preston, Nenad Miljkovic, Jean Sack, Ryan Enright, John Queeney, and
  Evelyn~N. Wang.
\newblock Effect of hydrocarbon adsorption on the wettability of rare earth
  oxide ceramics.
\newblock {\em Appl. Phys. Lett.}, 105(1), jul 2014.

\bibitem{Mamedov2023}
D.~Mamedov, A.~C. Åsland, S.~P. Cooil, H.~I. Røst, J.~Bakkelund,
  A.~Allaniyazov, J.~W. Wells, and S.~Karazhanov.
\newblock Enhanced hydrophobicity of {CeO}$_2$ thin films: {Role} of the
  morphology, adsorbed species and crystallography.
\newblock {\em Mater. Today Commun.}, 35:106323, June 2023.

\bibitem{Preisler2001}
E.~J. Preisler, O.~J. Marsh, R.~A. Beach, and T.~C. McGill.
\newblock Stability of cerium oxide on silicon studied by x-ray photoelectron
  spectroscopy.
\newblock {\em J. Vac. Sci. Technol., B: Microelectron. Nanometer Struct.
  Process., Meas., Phenom.}, 19(4):1611--1618, jul 2001.

\bibitem{Maslakov2018}
Konstantin~I. Maslakov, Yury~A. Teterin, Aleksej~J. Popel, Anton~Yu. Teterin,
  Kirill~E. Ivanov, Stepan~N. Kalmykov, Vladimir~G. Petrov, Peter~K. Petrov,
  and Ian Farnan.
\newblock {XPS} study of ion irradiated and unirradiated {CeO}$_2$ bulk and
  thin film samples.
\newblock {\em Appl. Surf. Sci.}, 448:154--162, August 2018.

\bibitem{Martinez2011}
L.~Martínez, E.~Román, J.~L. de~Segovia, S.~Poupard, J.~Creus, and
  F.~Pedraza.
\newblock Surface study of cerium oxide based coatings obtained by cathodic
  electrodeposition on zinc.
\newblock {\em Appl. Surf. Sci.}, 257(14):6202--6207, May 2011.

\bibitem{Khan2015}
Sami Khan, Gisele Azimi, Bilge Yildiz, and Kripa~K. Varanasi.
\newblock Role of surface oxygen-to-metal ratio on the wettability of
  rare-earth oxides.
\newblock {\em Appl. Phys. Lett.}, 106(6):061601, February 2015.

\bibitem{Burroughs1976}
Peter Burroughs, Andrew Hamnett, Anthony~F. Orchard, and Geoffrey Thornton.
\newblock Satellite structure in the {X}-ray photoelectron spectra of some
  binary and mixed oxides of lanthanum and cerium.
\newblock {\em J. Chem. Soc., Dalton Trans.}, (17):1686--1698, jan 1976.

\bibitem{Huefner2003}
Stefan Hüfner.
\newblock {\em Photoelectron Spectroscopy}.
\newblock Springer Berlin Heidelberg, 2003.

\bibitem{Astley2022}
Simon Astley, Di~Hu, Kerry Hazeldine, Johnathan Ash, Rachel E. Cross, Simon
  Cooil, Martin W. Allen, James Evans, Kelvin James, Federica Venturini, David
  C. Grinter, Pilar Ferrer, Rosa Arrigo, Georg Held, Gruffudd T. Williams,
  and D.~Andrew Evans.
\newblock Identifying chemical and physical changes in wide-gap semiconductors
  using real-time and near ambient-pressure {XPS}.
\newblock {\em Faraday Discuss.}, 236(0):191--204, 2022.

\bibitem{Sarapatka1992}
T.~J. Šarapatka.
\newblock {XPS} study of lead on {SiOx}/{Si}: temperature and light effects.
\newblock {\em Surf. Sci.}, 275(3):443--449, September 1992.

\bibitem{AcostaSilva2017}
Yuliana de~Jesus Acosta-Silva, Rebeca Castañedo-Perez, Gerardo Torres-Delgado,
  Arturo Méndez-López, and Orlando Zelaya-Ángel.
\newblock Effect of annealing temperature on structural, morphological and
  optical properties of {CeO}$_2$ thin films obtained from a simple precursor
  solution.
\newblock {\em J. Sol-Gel Sci. Technol.}, 82(1):20--27, April 2017.

\end{thebibliography}

\end{document}


\title[Water Absorption on $\mathrm{CeO}_{2}$ at Low Temperatures for Understanding Anti-Icing]{Supporting Information for Initial Stages of Water Absorption on $\mathbf{CeO}_{2}$ Surfaces at Very Low Temperatures for Understanding Anti-Icing Coatings}

\author{A C Åsland$^{1}$, S P Cooil$^{2}$, D Mamedov$^{3}$, H I Røst$^{1,4}$, J Bakkelund$^{1}$, Z Li$^{5}$, S Karazhanov$^{3}$ and J W Wells$^{1,2}$} 

\address{$^{1}$ Department of Physics, Norwegian University of Science and Technology (NTNU), NO-7491 Trondheim, Norway.}
\address{$^{2}$ Centre for Materials Science and Nanotechnology, University of Oslo, NO-0318 Oslo, Norway.}
\address{$^{3}$ Department for Solar Energy Materials and Technologies, Institute for Energy Technology, NO-2027 Kjeller, Norway.} 
\address{$^{4}$ Department of Physics and Technology, University of Bergen, NO-5007 Bergen, Norway.}
\address{$^{5}$ Department of Physics and Astronomy - Centre for Storage Ring Facilities (ISA), Aarhus University, DK-8000 Aarhus C, Denmark.} 

\ead{j.w.wells@fys.uio.no}
\vspace{10pt}
\begin{indented}
\item[]\today
\end{indented}

\section{Experimental Details}
\label{experimental}

\subsection{Scanning Electron Microscopy}
Scanning electron microscopy (SEM) was used to examine the surface morphology of the samples. Images were acquired utilising a JEOL model JSM-7900 F SEM. The image shown in Figure~1(a) in the main manuscript is of a $200\,$nm thick sample as grown.   
A similar surface morphology was observed also for samples with other $\mathrm{CeO}_{2}$ film thicknesses.

\subsection{X-ray Diffraction}
To characterise the crystal structure of the $\mathrm{CeO}_{2}$ coatings, X-ray diffraction (XRD) was measured using a Bruker D2 Phaser XRD with Cu-K$\alpha$ radiation. The measurements were conducted with Bragg-Brentano focusing and angles in the range $2\theta=20$-$60^{\circ}$. The samples were found to have a fluorite face-centred cubic lattice structure (space group 225). The measured lattice parameters of $a=5.415\,${\AA}~and $a=5.405\,${\AA}~for $200\,$nm and $400\,$nm thick coatings, respectively, agree with those found in literature \cite{Gerward2005}. 

\subsection{X-ray Photoelectron Spectroscopy}
The chemical composition and anti-icing properties of the $\mathrm{CeO}_{2}$ coatings were investigated by X-ray photoelectron spectroscopy (XPS). XPS measurements were performed at the AU-MatLine beamline on the ASTRID2 synchrotron facility at Aarhus University, using a SPECS PHOIBOS 150 electron analyser \cite{Xu2022}. 
The XPS measurements were acquired with photoexcitation energies of $h\nu=600\,$eV and $h\nu=650\,$eV. Pass energies were in the range $E_{P}=40$-$100\,$eV, and the energy resolution at the beamline was better than $0.1\,$eV. During the measurements, the pressure in the chamber was in the order of $10^{-10}\,$mbar.

\subsection{Calculating the Thickness of the Ice Overlayer from XPS}
The thickness of the ice formed on the surface of the cooled sample holder when it was exposed to an atmosphere of $\mathrm{H}_{2}\mathrm{O}$ was calculated from the XPS measurements using an overlayer method based on Beer-Lamberts law \cite{Sassaroli2004}. 
The intensity $I_{\mathrm{O}}$ of the XPS signal from an overlayer with thickness $t$ is proportional to the cross-section $\sigma_{\mathrm{O}}$, the atomic number density $N_{\mathrm{O}}$, and the attenuation of the electrons from a depth $z$, giving \cite{Song2012, Zemlyanov2018} 

\begin{equation}
\label{eq:IOverlayer}
    I_{\mathrm{O}}  = \sigma_{\mathrm{O}}N_{\mathrm{O}} \int_{0}^{t}\rme^{-z/\lambda_{\mathrm{O}}(E_{\mathrm{K}})} \rmd z = \sigma_{\mathrm{O}}N_{\mathrm{O}}\lambda_{\mathrm{O}}(E_{\mathrm{K}})\left[1-\rme^{-t/\lambda_{\mathrm{O}}(E_{\mathrm{K}})}\right] 
\end{equation}

where $\lambda_{\mathrm{O}}(E_{\mathrm{K}})$ is the inelastic mean-free-path of electrons in the overlayer with kinetic energy $E_{\mathrm{K}}$. 

Similarly, the intensity $I_{\mathrm{B}}$ of electrons from the bulk substrate is given by

\begin{equation}\label{eq:IBulk}
    I_{\mathrm{B}}=\sigma_{\mathrm{B}}N_{\mathrm{B}} \int_{t}^{\infty}\rme^{-z/\lambda_{\mathrm{B}}(E_{\mathrm{K}})} \rmd z = \sigma_{\mathrm{B}}N_{\mathrm{B}}\lambda_{\mathrm{B}}(E_{\mathrm{K}})\left[\rme^{-t/\lambda_{\mathrm{B}}(E_{\mathrm{K}})}\right].
\end{equation}

 Taking the ratio of equations \ref{eq:IOverlayer} and \ref{eq:IBulk}, and assuming that $\lambda_{\mathrm{O}}(E_{\mathrm{K}}) \approx \lambda_{\mathrm{B}}(E_{\mathrm{K}}) \approx \lambda$, one can estimate the thickness 

\begin{equation}
    t=\lambda\ln{\left( \frac{I_{\mathrm{O}}\sigma_{\mathrm{B}}N_{\mathrm{B}}}{I_{\mathrm{B}}\sigma_{\mathrm{O}}N_{\mathrm{O}}} + 1 \right)}
\end{equation}

of the overlayer. 

For the $\mathrm{H}_{2}\mathrm{O}$ overlayer on the sample holder, the intensity of the O~1s core level was used for $I_{\mathrm{O}}$, and the intensity of the Mo~3d core level was used for $I_{\mathrm{B}}$. The intensities were extracted from the overview spectra of the sample holder shown in Figure~2 in the main manuscript. Since there are few data points across the core levels in the overview spectra, there is a significant uncertainty in their intensity. This is in addition to uncertainties introduced by assuming a uniform overlayer and the same inelastic mean-free-path of electrons from different core levels and with different kinetic energies. 
The inelastic mean-free-path was approximated as $\lambda\approx0.5\,$nm since the kinetic energy of the electrons from the O~1s and Mo~3d core levels were in the range $60$-$380\,$eV \cite{Zangwill1988, Powell2020}.

\subsection{Normalising the Measured Spectra}

XPS measurements in a certain binding energy region were generally measured with the same photoexitiation energy, pass energy, and with the sample in the same position. Hence, intensities and backgrounds are expected to be comparable. 
During the analysis of the results, the measured spectra were compared to each other without normalisation, and after normalising by dividing by the total intensity or the total background intensity of each spectrum.   
When comparing spectra from the as-grown and air-treated samples before and after heating or exposing them to $\mathrm{H}_{2}\mathrm{O}$, the observed increase/decrease in the intensity of a given peak or component appear similar independently of the normalisation method used. 
The background of the O~1s spectra of the sample holder was significantly smaller than that of the as-grown and air-treated samples, and these spectra can therefore not be compared quantitatively. 

The overview spectra in Figures~1(c) and 2 in the main manuscript have been normalised by dividing by the total intensity of the relevant spectrum. 
In Figures~\ref{fig:C1s}, \ref{fig:O1s} and \ref{fig:Ce4p}, and in Figures~3 and 4 in the main manuscript, the total background intensity was used to normalise the spectra, and the background subtracted to make it easier to visually compare the peaks.  
The valence band spectra in Figure~\ref{fig:VB} was normalised by dividing by the average intensity in the region between $-0.5$-$1\,$eV where the background is nearly flat. 

When the relative intensity of different elements (e.g. Ce and O) or different peak components (e.g. $\mathrm{Ce}^{4+}$ and $\mathrm{Ce}^{3+}$) were calculated, the intensities of each peak or component were extracted from the same spectra. 
Since the relative intensities from a single spectrum are not affected by the overall intensity of the spectrum, these can be quantitatively compared for the different samples and sample treatments.

\section{Temperature Dependence of the Chemical Composition}

\begin{figure}
	\begin{center}
		\includegraphics[width=0.52\textwidth]{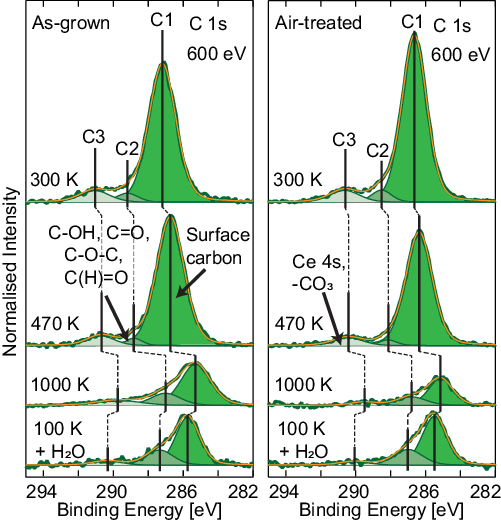}
		\caption{Photoemission intensity of the C~1s core level on the as-grown and air-treated samples measured before heating ($T=300\,$K), at $T\approx300\,$K after heating to $T=470\,$K and $T=1000\,$K, and whilst keeping the samples at $T=100\,$K after the $\mathrm{H}_{2}\mathrm{O}$ exposure. The peak labelled C1 (dark green) correspond to surface carbon, C2 (medium green) to $\mathrm{C-OH}$, $\mathrm{C=O}$, $\mathrm{C-O-C}$ and $\mathrm{C(H)=O}$, and C3 (light green) to $\mathrm{-CO}_{3}$ and the Ce 4s core level.}
		\label{fig:C1s}
	\end{center}
\end{figure}

\begin{figure}
	\begin{center}
		\includegraphics[width=0.52\textwidth]{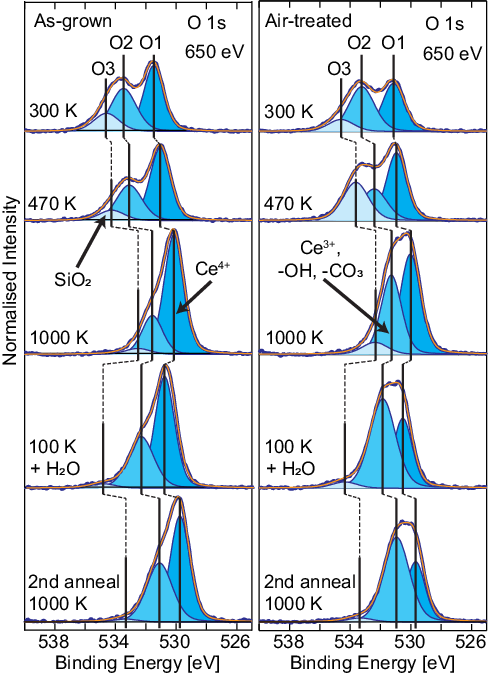}
		\caption{Photoemission intensity of the O~1s core level on the as-grown and air-treated samples measured before heating ($T=300\,$K), at $T\approx300\,$K after heating to $T=470\,$K and $T=1000\,$K, whilst keeping the samples at $T=100\,$K after exposure to $\mathrm{H}_{2}\mathrm{O}$, and at $T\approx300\,$K after heating the sample to $T=1000\,$K a second time after the $\mathrm{H}_{2}\mathrm{O}$ exposure. The O1 peak (dark blue) represents $\mathrm{Ce}^{4+}$, the O2 peak (medium blue) represents $\mathrm{Ce}^{3+}$, $\mathrm{-OH}$ and $\mathrm{-CO}_{3}$, and the O3 peak (light blue) represents $\mathrm{SiO}_2$.}
		\label{fig:O1s}
	\end{center}
\end{figure}

\begin{figure*}
	\begin{center}
		\includegraphics[width=\textwidth]{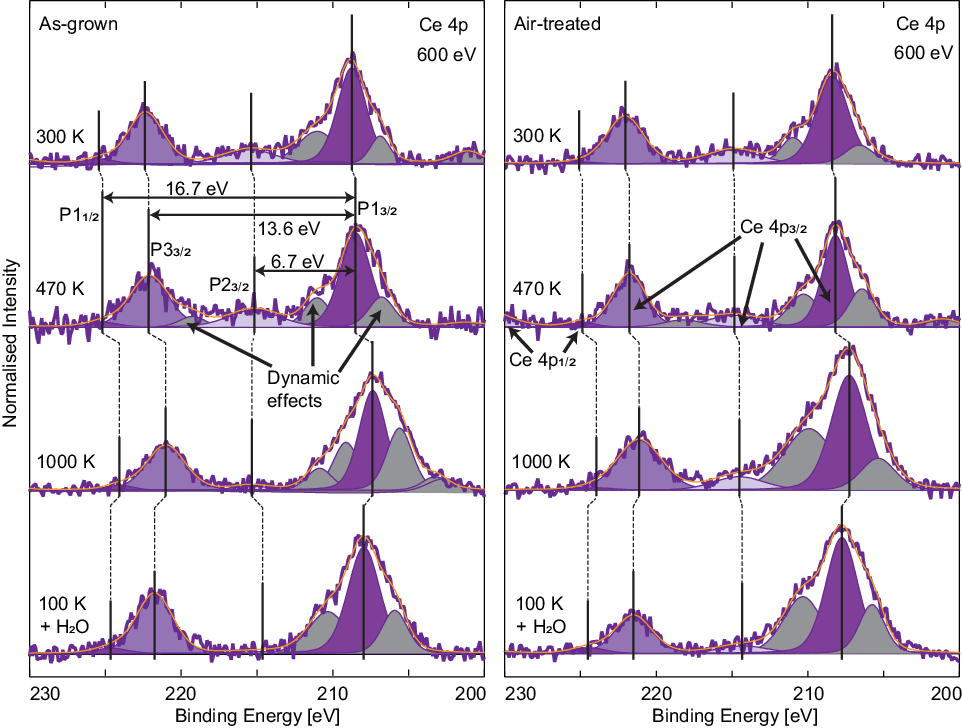}
		\caption{Photoemission intensity of the Ce~4p core level on the as-grown and air-treated samples measured before heating ($T=300\,$K), at $T\approx300\,$K after heating to $T=470\,$K and $T=1000\,$K, and whilst keeping the samples at $T=100\,$K after the $\mathrm{H}_{2}\mathrm{O}$ exposure. The P1 (dark purple), P2 (medium purple) and P3 (light purple) peaks represent different final states of Ce after the photoemission process. The peaks labelled dynamic effects (grey) are assigned to Coster-Kronig transitions. }
		\label{fig:Ce4p}
	\end{center}
\end{figure*}

Figures~\ref{fig:C1s}, \ref{fig:O1s} and \ref{fig:Ce4p} show how the chemical composition and structure of the samples evolved with temperature. The spectra labelled $300\,$K were measured just after introducing the samples into vacuum, but before heating. The spectra labelled $470\,$K and $1000\,$K were measured at $T\approx300\,$K after heating the samples briefly to $T=470\,$K and $T=1000\,$K, respectively. The spectra labelled $100\,\mathrm{K}+\mathrm{H}_{2}\mathrm{O}$ were measured whilst the samples were kept at $T\approx100\,$K, after exposure to $\mathrm{H}_{2}\mathrm{O}$ at $T\approx100\,$K.

Hydrocarbons on the surface of $\mathrm{CeO}_2$ have previously been shown to enhance and potentially be the origin of the hydrophobicity observed for $\mathrm{CeO}_{2}$ \cite{Preston2014, Mamedov2023}. 
The surface C -- therein the hydrocarbons, represented by the C1 (dark green) component in Figure~\ref{fig:C1s}, is significantly reduced upon heating, thus making it possible to investigate the ice formation or anti-icing properties of $\mathrm{CeO}_{2}$ without hydrocarbons on the surface. 
Since there are slightly more surface C on the air-treated sample compared to the as-grown sample at $T=300\,$K, but less after heating to $T>470\,$K, it seems like the surface C is more loosely bound if the sample has previously been heated in air.    
The C3 (light green) peak representing carbonates ($\mathrm{-CO}_{3}$) overlaps with the binding energy of the Ce 4s core level. Ce 4s is expected to increase with heating because more of the $\mathrm{CeO}_{2}$ is exposed when the surface C is removed. However, the intensity of C3 decrease upon heating and is therefore thought to be mainly from $\mathrm{-CO}_{3}$.
The amount of $\mathrm{C-OH}$, $\mathrm{C=O}$, $\mathrm{C-O-C}$ and $\mathrm{C(H)=O}$, represented by the C2 (medium green) peaks, does not change much with heating, but increases slightly after exposure to $\mathrm{H}_{2}\mathrm{O}$. It is possible that this is because the $\mathrm{H}_{2}\mathrm{O}$ reacts with the surface carbon to form such groups, or that more of these groups stick to the samples when they are cold. 

The evolution of the O~1s core level can be seen in Figure~\ref{fig:O1s}. The O1 (dark blue) component representing $\mathrm{Ce}^{4+}$ ($\mathrm{CeO}_{2}$) increases steadily upon heating for both samples, confirming that more of the $\mathrm{CeO}_{2}$ is exposed when the surface C is removed \cite{Preisler2001, Maslakov2018}. 
The O2 (medium blue) component has been ascribed to $\mathrm{Ce}^{3+}$, hydroxyls ($\mathrm{-OH}$) and carbonates ($\mathrm{-CO}_{3}$) \cite{Preisler2001, Maslakov2018, Martinez2011, Khan2015}. 
For both samples, the initial small decrease in the O2 intensity upon heating is thought to be due to the loss of $\mathrm{-CO}_{3}$ from the surface as seen in Figure~\ref{fig:C1s}. 
As discussed in the main manuscript, after heating to $1000\,$K, the majority of the O2 intensity seems to be from $\mathrm{Ce}^{3+}$ \cite{Preisler2001, Maslakov2018, Martinez2011, Khan2015}. 
The component labelled O3 (light blue) can be from $\mathrm{SiO}_2$ from the underlying oxidised Si substrate and from molecular $\mathrm{H}_{2}\mathrm{O}$ \cite{Preisler2001, Martinez2011, Khan2015}. However, since $\mathrm{H}_{2}\mathrm{O}$ evaporates at $T<370\,$K for pressures below 1\,atm, the intensity of O3 is thought to be from $\mathrm{SiO}_2$, and the intensity variations related to the uniformity of the sample and how oxidised the Si substrate is.  

The XPS measurements of the Ce~4p core level seen in Figure~\ref{fig:Ce4p} show only minor changes when heating and cooling and exposing the samples to $\mathrm{H}_{2}\mathrm{O}$.    
A more extensive discussion of the spectra from Ce can be found in the main manuscript and References \cite{Maslakov2018} and \cite{Burroughs1976}. 
It should be noted that the total intensity of the Ce~4p spectra increases after the samples have been heated, which is as expected when surface C is removed. 
Additionally, the intensity of particularly the P2 peak (light purple) vary with both temperature and sample. Hence, which final state is favourable seems to depend on the heating temperature and whether the sample has previously been heated in air. 

\section{Energy Shifts in the XPS Measurements}

\begin{figure}
	\begin{center}
		\includegraphics[width=0.5\textwidth]{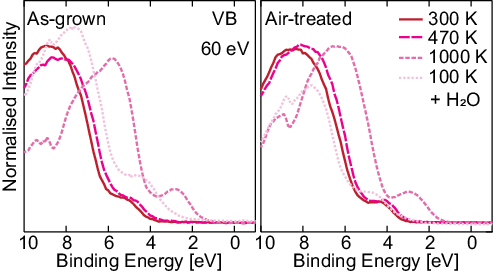}
		\caption{Photoemission intensity of the valence band of the as-grown and air-treated samples measured before heating ($T=300\,$K), at $T\approx300\,$K after heating to $T=470\,$K and $T=1000\,$K and whilst keeping the samples at $T=100\,$K after the $\mathrm{H}_{2}\mathrm{O}$ exposure.}
		\label{fig:VB}
	\end{center}
\end{figure}

The solid and dashed vertical lines in Figures~\ref{fig:C1s}, \ref{fig:O1s} and \ref{fig:Ce4p} indicate the peak positions of the main peaks in the XPS measurements. The peaks generally shift towards lower binding energies upon heating, but shift (partly) back towards higher binding energies upon cooling and $\mathrm{H}_{2}\mathrm{O}$ exposure. This is also clear from the valence band (VB) spectra shown i Figure~\ref{fig:VB}.  
Shifts of the O~1s core level towards higher binding energies upon heating or irradiation by ions in vacuum have been observed previously as a consequence of $\mathrm{Ce}^{4+}$ reducing to $\mathrm{Ce}^{3+}$, however, there the shifts are in the opposite direction compared to these samples \cite{Preisler2001, Maslakov2018}. Hence, the shifts observed in our measurements seem to be caused by something else.   

The binding energies are very sensitive to the chemical structure and the electrostatic potential on the surface of the measured sample \cite{Huefner2003, Astley2022}.  
Species adsorbed on the surface can act as dopants, causing changes in the surface potential and result in binding energy shifts \cite{Astley2022}. 
After cooling and $\mathrm{H}_{2}\mathrm{O}$ exposure, the VB spectra of the as-grown sample only shifted partly back towards the binding energy it had before heating. This can indicate that some of the surface contaminants were acting as $n$-type dopants before heating, but were permanently removed in the heating process. For the sample which was heated in air after growth, however, the binding energy of the VB maxima before heating and after cooling and depositing $\mathrm{H}_{2}\mathrm{O}$ are similar to each other, and also similar to that of the as-grown sample after heating. This indicates that adsorbed species are removed or the chemical structure on the surface changed the first time a sample is heated after growth, resulting in a binding energy shift towards lower binding energies. 

For semiconductors, the charge distribution at the surface or at the interface between different semiconductors can lead to band bending, causing binding energy shifts in the measured spectra \cite{Astley2022}.  
The energy gap and Fermi-level of a semiconductor is often temperature dependent -- and consequently, the band bending and therefore the binding energy shift are also temperature dependent \cite{Sarapatka1992}.
Changes in the the band bending at the $\mathrm{CeO}_{2}$ surface or at the interface between $\mathrm{CeO}_{2}$ and the Si substrate is therefore a possible explanation of the binding energy shift. 

If the band gap is very large, a large resistivity in the bulk can delay charge transfer from the bulk to the surface, causing surface charging and binding energy shifts \cite{Astley2022}. In this case a larger resistivity will lead to more charging and give a binding energy shift towards larger binding energies \cite{Astley2022}. Furthermore, if the resistivity is temperature dependent, the binding energy shift will be temperature dependent too. 
The VB spectra in Figure~\ref{fig:VB} shift towards lower binding energies after heating, suggesting that the resistivity of $\mathrm{CeO}_{2}$ decreases with temperature if the shift is caused by surface charging. This is in agreement with a previous study which shows that the electrical resistivity of $\mathrm{CeO}_{2}$ decreases with temperature, and can thus be another explanation of the binding energy shift \cite{AcostaSilva2017}. 

Hence, there are several possible explanations of the observed binding energy shifts related to the details of the chemical structure and electrostatic potential at the surface.

\section*{References}

\bibliographystyle{unsrt}
\bibliography{CeO2TempDep}